\documentclass{appolb}
\usepackage{graphicx}
\usepackage{xcolor}
\usepackage{amsmath, empheq, amsfonts, amssymb}

\begin{document}
\title{Dynamical Hadrons%
\thanks{Presented by Vin\'icius Rodrigues Debastiani at Excited QCD 2018 - Kapaonik, Serbia. 11-15 March 2018.}%
    }
\author{V.~R.~Debastiani$^1$,\\
 E.~Oset$^1$, J.~M.~Dias$^2$, Wei-Hong~Liang$^3$\\
    \address{$^1$Departamento de
    F\'{\i}sica Te\'orica and IFIC, Centro Mixto Universidad de
    Valencia-CSIC, Institutos de Investigaci\'on de Paterna, Aptdo.
    22085, 46071 Valencia, Spain\\
    $^2$Instituto de F\'{i}sica, Universidade de S\~{a}o Paulo, Rua do Mat\~{a}o, 1371, Butant\~{a}, CEP 05508-090, S\~{a}o Paulo, S\~{a}o Paulo, Brazil\\
    $^3$Department of Physics, Guangxi Normal University,
    Guilin 541004, China\\
    }
 }

\maketitle

\begin{abstract}
In this talk I briefly review the theory of resonances dynamically generated from hadron-hadron scattering, sometimes referred to as ``molecules''. I give some classical examples of meson-meson and meson-baryon systems, as well as a few examples of different approaches to describe the interaction between hadrons. To conclude, I comment on a few recent works that suggest that some of the five new narrow $\Omega_c$ states, recently discovered by the LHCb collaboration, can be interpreted as meson-baryon molecules.
\end{abstract}

\section{Introduction}

There are many states that can be described from hadron-hadron interaction. Some well-known examples are the scalar mesons obtained from pseudoscalar-pseudoscalar interaction in $S$-wave and coupled channels: the
  $a_0(980)$ from $K \bar K$ and $\pi \eta$ in isospin 1,
  the $f_0(980)$ from $K \bar K$ and $\pi \pi$ in isospin 0,
  and the $f_0(500)$ ($\sigma$ meson) from $\pi\pi$ scattering in isospin 0.
 In the strange sector, from vector-pseudoscalar interaction one can describe the $f_1(1285)$ as a $K^\ast \bar K + c.c.$ molecule. In the charm-strange sector there is the $D_{s0}^\ast(2317)$, which can be described as a $DK$ bound state. Similarly, one of the most famous examples in charm sector is the $X(3872)$ which can be explained as a $D \bar D^{\ast} + c.c.$ molecule.

These are just a few cases from meson-meson interaction. On the other hand, in meson-baryon interaction the best example would be
the $\Lambda(1405)$, which is widely accepted \cite{LambdaPDG} as a quasi-bound state between the $\bar K N$ and $\pi \Sigma$ thresholds, generated mostly from the $\bar K N$ scattering.
In the charm sector, the $\Lambda_c(2595)$ shows a similar pattern, lying between the $DN$ and $\pi \Sigma_c$ thresholds.
Another good example is the $N^\ast(1535)$, which is also well described from $\eta N$ and $\pi N$ interaction. In this context, one could wonder if the new $\Omega_c$ states \cite{OmegacLHCb} could be described from the interaction of $\Xi_c \bar K$ and its coupled channels.

\section{Chiral Unitary approach and the Local Hidden Gauge}

Assuming the on-shell factorization of the Bethe-Salpeter equation, we can obtain a unitarized scattering amplitude, using an effective interaction $V$ where the hadrons are the degrees of freedom,
\begin{equation}
\label{bs}
 T(s) = [1 - V(s) G(s)]^{-1}V(s),
\end{equation}
with $G$ the meson-meson or meson-baryon loop function. Then we can look for poles of the amplitude in the complex energy plane, which are related with the mass and width of the resonances by $z_R = M_R - i \Gamma_R/2$. %Poles above threshold in the second Riemann sheet are interpreted as resonances, whereas poles below threshold on the real axis of the first sheet are interpreted as bound states.

The meson-baryon interaction in the $\textrm{SU(3)}$ sector can be described by the chiral Lagrangian
\begin{eqnarray}
\label{lag}
\nonumber \mathcal{L}^B = \frac{1}{4f_{\pi}^2}\,  \left\langle \bar{B} i \gamma^{\mu} \Big[ (\Phi\, \partial_{\mu}\Phi -  \partial_{\mu}\Phi\,\Phi\,)B -
B(\Phi\, \partial_{\mu}\Phi -  \partial_{\mu}\Phi\,\Phi\,) \Big ] \right\rangle \, ,
\end{eqnarray}

\begin{equation}
\label{phimatrix}
 \Phi =
\left(
\begin{array}{ccc}
\frac{1}{\sqrt{2}} \pi^0 + \frac{1}{\sqrt{6}} \eta & \pi^+ & K^+ \\
\pi^- & - \frac{1}{\sqrt{2}} \pi^0 + \frac{1}{\sqrt{6}} \eta & K^0 \\
K^- & \bar{K}^0 & - \frac{2}{\sqrt{6}} \eta
\end{array}
\right)\, ,
\end{equation}
%\quad
\begin{equation}
\label{Bmatrix}
\nonumber B =
\left(
\begin{array}{ccc}
\frac{1}{\sqrt{2}} \Sigma^0 + \frac{1}{\sqrt{6}} \Lambda &
\Sigma^+ & p \\
\Sigma^- & - \frac{1}{\sqrt{2}} \Sigma^0 + \frac{1}{\sqrt{6}} \Lambda & n \\
\Xi^- & \Xi^0 & - \frac{2}{\sqrt{6}} \Lambda
\end{array}
\right) \, .
\end{equation}

 At energies close to threshold one can consider only the dominant contribution coming from $\partial_0$ and $\gamma^0$, such that the interaction is given by
 \vspace{-5pt}
\begin{eqnarray}
\label{kernel}
V_{ij}= -C_{ij} \frac{1}{4f_\pi^2}(k^0+k^{\prime 0})\, ,
\end{eqnarray}
%\vspace{-5pt}
where $k^0$ and $k^{\prime 0}$ are the energies of the incoming and outgoing mesons, respectively.
This framework was used to described the $\Lambda(1405)$ with coupled channels in $\bar K N$ scattering \cite{Lambda}.

%\begin{figure}
%	\begin{center}
%		\includegraphics[width=0.7\textwidth]{diagKp.eps}
%	\end{center}
%	\caption{\label{diagKp} Vector exchange in the meson-baryon interaction describing the $\Lambda(1405)$.}
%\end{figure}

Alternatively, one can use the Local Hidden Gauge Approach (LHGA), where the meson-baryon interaction in $\textrm{SU(3)}$ is obtained exchanging vector mesons. %, as depicted in Fig.~\ref{diagKp}.
The ingredients needed are the $VPP$ and $VBB$ Lagrangians
\vspace{-10pt}
\begin{eqnarray}
\mathcal{L}_{VPP} = -i g\, \langle \,\, [\,\Phi,\,\partial_{\mu} \Phi\,]\, V^{\mu}\, \rangle \, ,
\end{eqnarray}
\vspace{-20pt}
\begin{eqnarray}
\label{vbbLag}
&\mathcal{L}_{VBB} = g\, \Big( \langle \bar{B} \gamma_{\mu} [V^{\mu},B] \,\rangle +
\langle \bar{B} \gamma_{\mu} B \rangle \langle V^{\mu} \rangle \Big) \, ,
\end{eqnarray}
\vspace{-5pt}
where
\vspace{-5pt}
\begin{equation}
\label{vfields}
 V^\mu=\left(
\begin{array}{ccc}
\frac{\rho^0}{\sqrt{2}}+\frac{\omega}{\sqrt{2}} & \rho^+ & K^{*+}\\
\rho^- &-\frac{\rho^0}{\sqrt{2}}+\frac{\omega}{\sqrt{2}} & K^{*0}\\
K^{*-} & \bar{K}^{*0} &\phi\\
\end{array}
\right)^\mu\ ,
\end{equation}
and $g=m_V/2f_\pi$, with $m_V$ the mass of the vector mesons ($\sim 800$ MeV).

Taking $q^2/m^2_V\to 0$ in the propagator of the exchanged vector
gives rise to the same interaction of the Chiral Lagrangian.

%\subsection{Molecular description of the narrow $\Omega_c$ states}

In our work \cite{Omegac} we have studied meson-baryon molecular states with $C=+1$, $S=-2$, $I=0$ to investigate if some of the recently discovered $\Omega_c$ states \cite{OmegacLHCb} can be described as dynamically generated resonances.

Inspired in the Lagrangians of the LHGA, the meson-baryon interaction was described through the exchange of vector mesons.
Extending the $VPP$ Lagrangian to the charm sector is simple. We take the same structure including a fourth line and column with the charmed pseudoscalars in $\Phi$ and with the charmed vector mesons in $V^\mu$.
%\vspace{-5pt}
%\begin{equation}
%\label{Pmatrix}
% P =
%\left(
%\begin{array}{cccc}
%\frac{1}{\sqrt{2}} \pi^0 + \frac{1}{\sqrt{3}} \eta  + \frac{1}{\sqrt{6}} \eta^{\prime} & \pi^+ & K^+ & \bar{D}^0 \\
%\pi^- &  - \frac{1}{\sqrt{2}} \pi^0 +\frac{1}{\sqrt{3}} \eta + \frac{1}{\sqrt{6}} \eta^{\prime} & K^0 & D^- \\
%K^- & \bar{K}^0 & -\frac{1}{\sqrt{3}} \eta +\sqrt{\frac{2}{3}} \eta^{\prime} & D_s^- \\
%D^0 & D^+ & D_s^+ & \eta_c\\
%\end{array}
%\right)\, ,
%\end{equation}
%where we also include the mixing between $\eta$ and $\eta^{\prime}$, and
%\begin{equation}
%\label{Vmatrix}
% V =
%\left(
%\begin{array}{cccc}
%\frac{1}{\sqrt{2}} \rho^0 + \frac{1}{\sqrt{2}} \omega & \rho^+ & K^{* +} & \bar{D}^{* 0} \\
%\rho^- & -\frac{1}{\sqrt{2}} \rho^0 + \frac{1}{\sqrt{2}} \omega & K^{* 0} & \bar{D}^{* -} \\
%K^{* -} & \bar{K}^{* 0} & \phi & D_s^{* -} \\
%D^{* 0} & D^{* +} & D_s^{* +} & J/\psi\\
%\end{array}
%\right)\, .
%\end{equation}

Extending the $VBB$ Lagrangian to the charm sector is not so easy.
Instead of using $SU(4)$ symmetry, we look at the quark structure of the exchanged vectors like $\rho^0$, $\omega$ and $\phi$ (which can be extended to $K^*$, $\rho^{\pm}$, etc)
\vspace{-2pt}
\begin{eqnarray}
\label{waves}
\rho^0 &=& (u\bar{u} - d\bar{d})/\sqrt{2}\, ,\nonumber \\
\omega &=& (u\bar{u} + d\bar{d})/ \sqrt{2}\, , \\
\phi &=& s\bar{s} \, .\nonumber
\end{eqnarray}
Within the approximation of $\gamma^{\mu} \to \gamma^0$ we have no spin dependence in the baryon sector related to the exchanged vector, and we can consider an operator at the quark level.
This way, applying the vector meson as a number operator in the wave functions of the baryons, we can get the same result as using $\mathcal{L}_{VBB}$ in $SU(3)$.

With that in mind, we have built spin-flavor wave functions for the baryons involved in our calculations, considering the heavy quark as spectator and using $SU(3)$ symmetry in the light quarks.
For instance, for the $\Xi^+_c$ we have a wave function antisymmetric for the two light quarks, both in flavor and spin
\vspace{-2pt}
\begin{equation}
\Xi^+_c = c(us - su) \uparrow(\uparrow \downarrow - \downarrow \uparrow)/2
\end{equation}
%\vspace{-2pt}
%While for the $\Xi^{\prime\,+}_c$ it is symmetric for the two light quarks
%\vspace{-2pt}
%\begin{equation}
%  \Xi^{\prime\,+}_c = c(us+su)[\uparrow(\uparrow\downarrow + \downarrow\uparrow) - 2 \downarrow\uparrow\uparrow)/(2\sqrt{3}),
%\end{equation}
%%\vspace{-5pt}
%And for $\Xi^{*+}_c$ the flavor part for the two light quarks is symmetric and all the spins are aligned
%\vspace{-7pt}
%\begin{equation}
%  \Xi^{*+}_c = c(us+su)\uparrow\uparrow\uparrow/\sqrt{2}
%\end{equation}

\begin{figure}[h!]
	\begin{center}
		\includegraphics[width=1.0\textwidth]{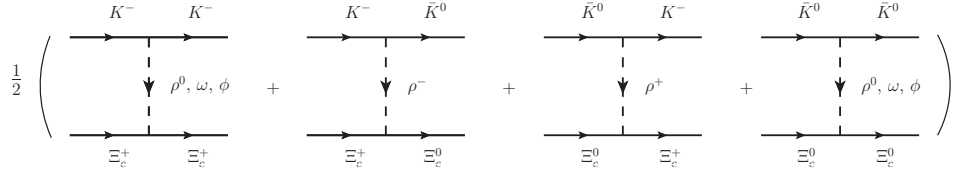}
	\end{center}
	\caption{\label{diag1App} Diagrams in the $\Xi_c\bar{K} \to \Xi_c\bar{K}$ interaction through vector meson exchange.}
\end{figure}

%This way we reach the same structure for the meson-baryon interaction and the
Finally, the coefficients of the $V_{ij}$ matrix can be constructed from the $VPP$ and $VBB$ vertex, taking into account all the diagrams as in Fig.~\ref{diag1App}.

Assuming the heavy quark is a spectator, implies that the interactions are dominated by the $SU(3)$ content of $SU(4)$. This has as a consequence that heavy quark spin symmetry is respected, except for a few nondiagonal transitions like $\Xi_c \bar{K}\to \Xi D$, where one has to exchange a $D^*_s$. In this case $SU(4)$ is used. However, these terms are suppressed by the heavy quark propagator that goes like $(1/m_{D^*_s})^2$.

In Tables~\ref{res1} and \ref{res4} we show the poles found from the interaction of pseudoscalar($0^-$)-baryon($1/2^+$) and pseudoscalar($0^-$)-baryon($3/2^+$), respectively. We also show the coupling $g_i$ of each pole to the channels it couples in our framework, and the quantity $g_iG_i^{II}$, which is proportional to the strength of the wave function at the origin (for $S$-wave) and is related to the strength of that channel to produce the resonance. In Table~\ref{tab:tab1} we compare our results with the measurements of the LHCb Collaboration \cite{OmegacLHCb}.
\vspace{-6pt}
\begin{table}[h!]
\caption{The coupling constants to various channels for the poles in the $J^P={1/2}^-$ sector, with $q_{max}=650$ MeV. $g_i\,G^{II}_i$ is in MeV.\label{res1}}
\centering  \footnotesize
\begin{tabular}{c c c c c}
\hline\hline
\noalign{\vskip2pt}
${\bf 3054.05+i0.44}$ & $\Xi_c\bar{K}$ & $\Xi^{\prime}_c\bar{K}$ & $\Xi D$ & $\Omega_c \eta$ \\
\hline
\noalign{\vskip1pt}
$g_i$ & $-0.06+i0.14$ & $\bf 1.94+i0.01$ & $-2.14+i0.26$ & $1.98+i0.01$ \\
$g_i\,G^{II}_i$ & $-1.40-i3.85$ & $\bf -34.41-i0.30$ &  $9.33-i1.10$ &  $-16.81-i0.11$ \\
\hline\hline
\noalign{\vskip2pt}
${\bf 3091.28+i5.12}$ & $\Xi_c\bar{K}$ & $\Xi^{\prime}_c\bar{K}$ & $\Xi D$ & $\Omega_c \eta$ \\
\hline
\noalign{\vskip1pt}
$g_i$ & $0.18-i0.37$ & $0.31+i0.25$ & $\bf 5.83-i0.20$ & $0.38+i0.23$ \\
$g_i\,G^{II}_i$ & $5.05+i10.19$ & $-9.97-i3.67$ & $\bf-29.82+i0.31$ & $-3.59-i2.23$ \\
\hline\hline
\end{tabular}
\end{table}

We see that from the pseudoscalar($0^-$)-baryon($1/2^+)$ interaction the mass and width of the $\Omega_c(3050)$ and $\Omega_c(3090)$ can be obtained with remarkable agreement with experiment. This implies that both of these states have quantum numbers $J^P=1/2^-$, and the couplings and wave functions tell us that the $\Omega_c(3050)$ is mostly a $\Xi^{\prime}_c\bar{K}$ molecule, which also couples strongly to the channels $\Xi D$ and $\Omega_c \eta$. The small coupling of the $\Omega_c(3050)$ to $\Xi_c \bar{K}$, the only channel open for strong decay (with a small phase space available), explains its extremely small width. On the other hand, the $\Omega_c(3090)$ is mostly a $\Xi D$ molecule, and the small width can also be explained by the small couplings to $\Xi_c \bar{K}$ and $\Xi^\prime_c K$, the former being slightly bigger and with more phase space than for the $\Omega_c(3050)$.
%\vspace{-5pt}
\begin{table}[h!]
\caption{The coupling constants to various channels for the poles in the $J^P={3/2}^-$ sector, with $q_{max}=650$ MeV. $g_i\,G^{II}_i$ is in MeV.\label{res4}}
\centering \footnotesize
\begin{tabular}{c c c c}
\hline\hline
\noalign{\vskip2pt}
${\bf 3124.84}$ & $\Xi^*_c \bar{K}$ ~&~ $\Omega^*_c \eta$ ~&~ $\Xi^* D$ \\
\hline
\noalign{\vskip1pt}
$g_i$ & $\bf 1.95$ &  $1.98$ & $-0.65$ \\
$g_i\,G^{II}_i$ &  $\bf -35.65$ & $-16.83$ & $1.93$ \\
\hline\hline
\noalign{\vskip2pt}
$3290.31+i0.03$ & $\Xi^*_c \bar{K}$ ~&~ $\Omega^*_c \eta$ ~&~ $\Xi^* D$ \\
\hline
\noalign{\vskip1pt}
$g_i$ & $0.01+i0.02$ & $0.31+i0.01$  & $\bf 6.22-i0.04$\\
$g_i\,G^{II}_i$ & $-0.62-i0.18$ & $-5.25-i0.18$ & $\bf-31.08+i0.20$\\
\hline\hline
\end{tabular}
\end{table}

The states  from pseudoscalar($0^-$)-baryon($3/2^+)$ interaction present a similar pattern, with a state corresponding to the $\Omega_c(3119)$ made mostly of $\Xi^{\ast}_c\bar{K}$. %(note that the $\Xi^{\ast}_c$ has the same flavor structure of the $\Xi^\prime_c$).
The decay of this state into $\Xi_c \bar{K}$, as seen in the experiment \cite{OmegacLHCb}, requires the exchange of vector mesons in $P$-wave, which give raise to its small width.
Another state made mostly of $\Xi^{\ast} D$ is also found, as well as others from the interaction of vector mesons with baryons, which are presented and discussed in Ref.~\cite{Omegac}.
\vspace{-6pt}
\begin{table}[h!]
\centering \footnotesize
  \caption{Comparison of our results \cite{Omegac} with the LHCb data \cite{OmegacLHCb}.\label{tab:tab1}}
\begin{tabular}{lccc}
    \hline\hline
\noalign{\vskip2pt}
Resonance  &  Mass (MeV) & $\Gamma$ (MeV) \cr
\hline
\noalign{\vskip2pt}
$\Omega_c(3050)$ & $3050.2 \pm 0.1 \pm 0.1 ^{+0.3}_{-0.5}$ & $0.8 \pm 0.2 \pm 0.1$ & $< 1.2, \; 95\%$ {\rm CL}  \cr
\noalign{\vskip2pt}
\noalign{\vskip2pt}
    This work \cite{Omegac}                 & $\bf  3054.05$  & $ \bf 0.88$ & \cr
\hline
\noalign{\vskip2pt}
$\Omega_c(3090)$ & $3090.2 \pm 0.3 \pm 0.5 ^{+0.3}_{-0.5}$ & $8.7 \pm 1.0\pm 0.8 $ & \cr
\noalign{\vskip2pt}
    This work \cite{Omegac}                 & ${\bf 3091.28}$ & $\bf 10.24$ &  \cr
\noalign{\vskip2pt}
    \hline
\noalign{\vskip2pt}
$\Omega_c(3119)$ & $3119.1 \pm 0.3 \pm 0.9 ^{+0.3}_{-0.5}$ & $1.1 \pm 0.8\pm 0.4 $ & $< 2.6 \;{\rm MeV}, 95\%$ {\rm CL}  \cr
\noalign{\vskip2pt}
\noalign{\vskip2pt}
  This work \cite{Omegac}                   & ${\bf  3124.84}$       & $\bf  0$ & \cr
\noalign{\vskip2pt}
\hline\hline
  \end{tabular}
\end{table}

%\vspace{-3pt}
The $\Omega_b^-$ baryons have already been measured, most recently by the LHCb Collaboration \cite{OmegabLHCb}. The search for the new $\Omega_c^0$ states on the weak decay of the $\Omega_b^-$ was recently proposed in Ref.~\cite{Omegab-Omegac}. Using our molecular description in coupled channels we  presented predictions for the reactions: $\Omega_b^- \to (\Xi \, D) \, \pi^-$, $\Omega_b^- \to (\Xi_c \, \bar{K}) \, \pi^-$ and   $\Omega_b^- \to (\Xi_c^\prime \, \bar{K}) \, \pi^-$.

The transition $\Xi D\to\Xi_c \bar K$ appears naturally in the coupled channels approach, and we expect to see the $\Omega_c(3050)$ and $\Omega_c(3090)$ in the $\Xi_c \bar K$ invariant mass distribution. As discussed in detail in Ref.~\cite{Omegab-Omegac}, the amplitude of the reaction $\Omega_b^- \to (\Xi_c \, \bar{K}) \, \pi^-$ would be of the form
%\vspace{-3pt}
\begin{equation}\label{tKbarXic}
t_{\Omega_b^- \to \pi^-  \Xi_c \bar K} = V_P \,%\frac{2}{\sqrt{3}}
G_{\Xi D}[M_{\rm inv}(\Xi_c \bar K)] \, t_{\Xi D\to\Xi_c \bar K}[M_{\rm inv}(\Xi_c \bar K)].
\end{equation}

\begin{figure}[h!]
  \centering
  \includegraphics[width=0.49\textwidth]{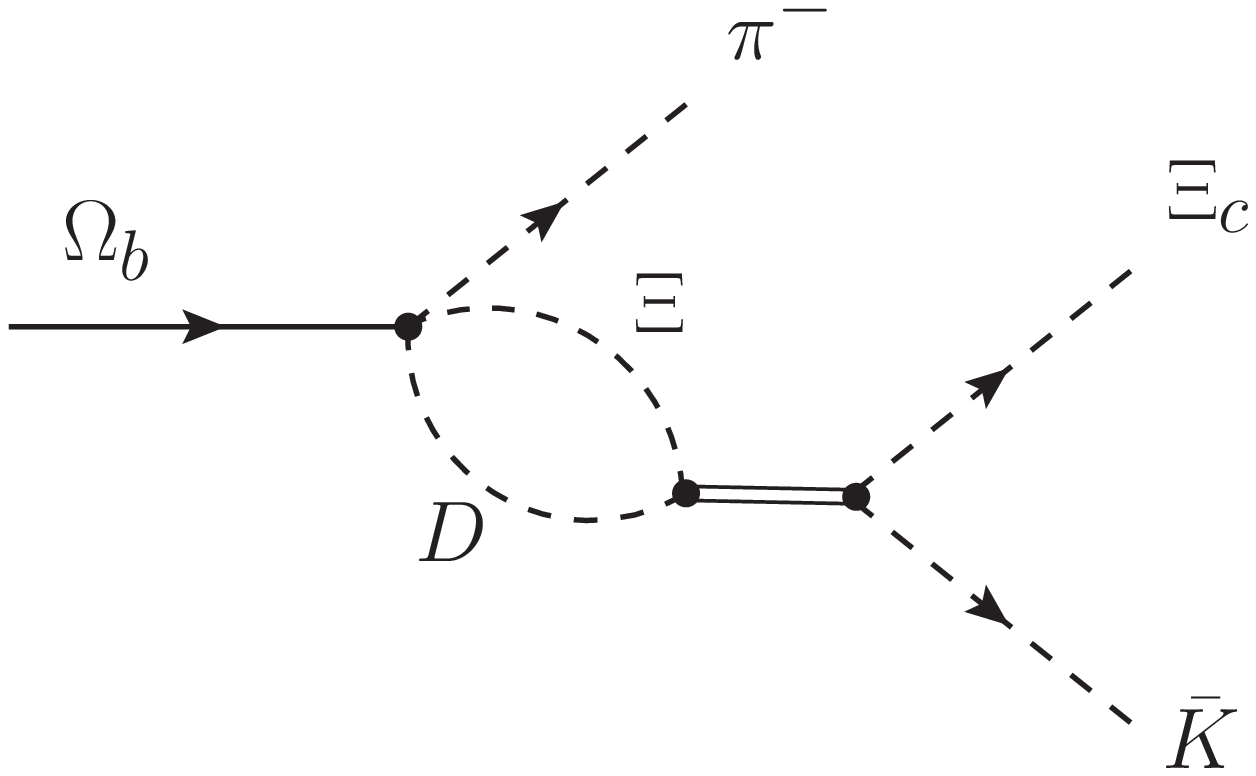}
  \includegraphics[width=0.5\textwidth]{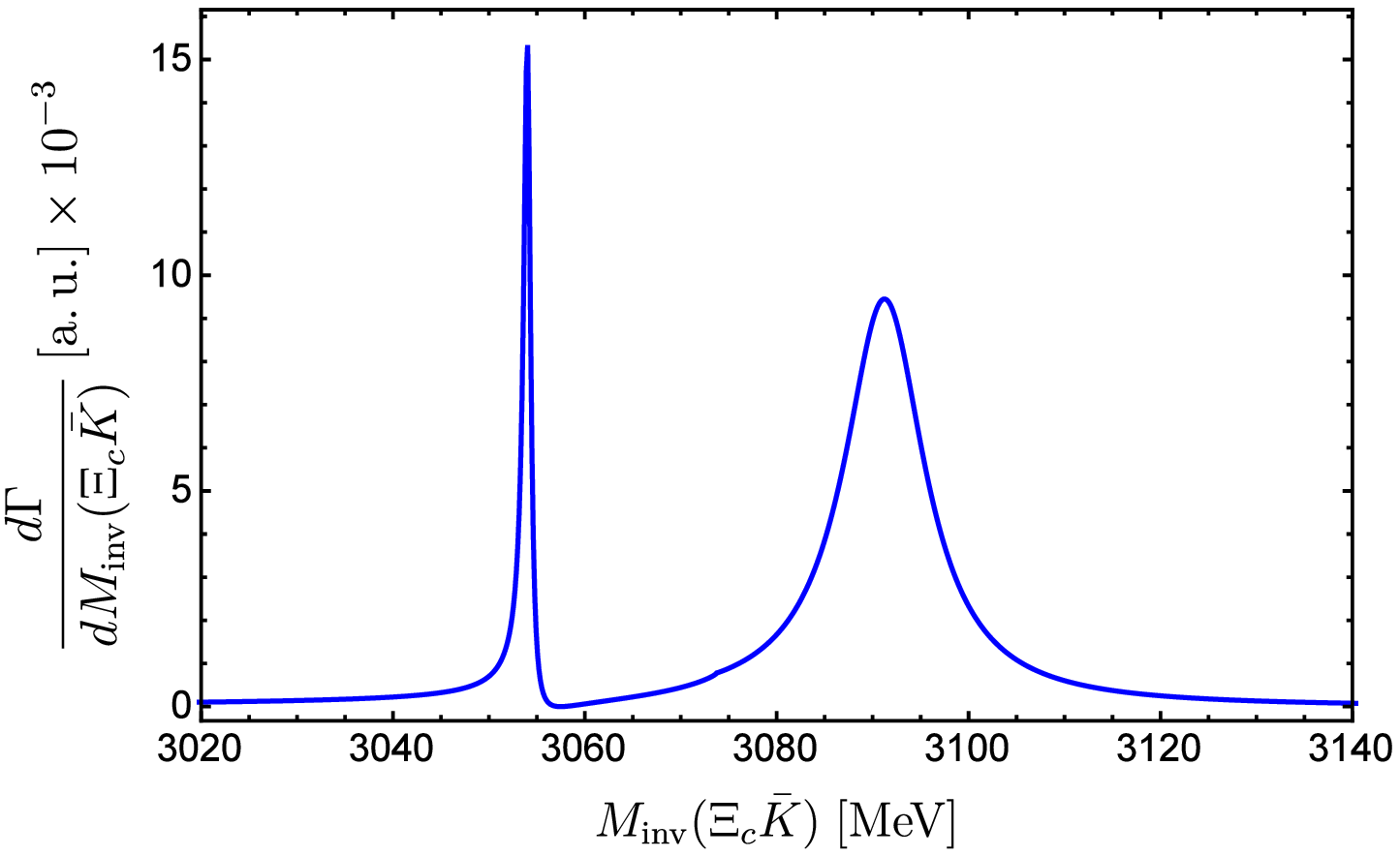}
\caption{$\Omega_b^- \to \pi^- \Xi_c \bar{K}$ process through $\Xi D$ rescattering. On the right, $\Xi_c \bar{K}$ invariant mass distribution showing the $\Omega_c(3050)$ and $\Omega_c(3090)$. \label{fig:Omegab-Omegac}}
\end{figure}

%\vspace{-3pt}
%Using our coupled channels approach from Ref.~\cite{Omegac}
Then we can estimate the  invariant mass distribution, as seen in Fig.~\ref{fig:Omegab-Omegac}.
And we can also predict the following rate of production
%\vspace{-3pt}
\begin{equation}
    \frac{\Gamma_{\Omega_b^- \to \pi^- {\mathbf{\Omega_c(3050)}}}}{\Gamma_{\Omega_b^- \to \pi^- {\mathbf{\Omega_c(3090)}}}} \approx 10\%.
\end{equation}
%\vspace{-3pt}

\vspace{-5pt}
\section{Acknowledgements}
%\vspace{-2pt}

V.~R.~Debastiani would like to thank the invitation to present this talk, and the organizers of the event for providing such a nice environment, specially Pedro Bicudo, Marina Marinkovic and Robert Kaminski for their extra effort in making this event possible.

V. R. D. also acknowledges the Programa Santiago Grisolia of Generalitat Valenciana (Exp. GRISOLIA/2015/005). J. M. D. thanks the Funda\c c\~ao de Amparo \`a Pesquisa do Estado de S\~ao Paulo (FAPESP) for support by FAPESP grant 2016/22561-2.
This work is partly supported by the Spanish Ministerio de Economia y Competitividad and European FEDER funds under the contract number FIS2014-57026-REDT, FIS2014-51948-C2- 1-P, and FIS2014-51948-C2-2-P, and the Generalitat Valenciana in the program Prometeo II-2014/068 (E.O.); and by the National Natural Science Foundation of China under Grants No. 11565007, No. 11747307 and No. 11647309.% (W.-H.~L.).

\vspace{-5pt}
\bibliographystyle{plain}

\begin{thebibliography}{99}
%\vspace{-3pt}
\vspace{-3pt}
\bibitem{LambdaPDG}
  Ulf-G. Mei{\ss}ner, T. Hyodo, \textit{Pole structure of the $\Lambda(1405)$ region}, in C. Patrignani {\it et al.} (Particle
Data Group), Chin. Phys. C {\bf 40}, 100001 (2016).

\bibitem{OmegacLHCb}
  R.~Aaij {\it et al.} [LHCb Collaboration],
  %``Observation of five new narrow $\Omega_c^0$ states decaying to $\Xi_c^+ K^-$,''\\
  Phys.\ Rev.\ Lett.\  {\bf 118}, 182001 (2017).

\bibitem{Lambda}
   E.~Oset and A.~Ramos,
  %``Nonperturbative chiral approach to s wave anti-K N interactions,''
  Nucl.\ Phys.\ A {\bf 635}, 99 (1998).

\bibitem{Omegac}
  V.~R.~Debastiani, J.~M.~Dias, W.~H.~Liang and E.~Oset,
  %``Molecular $\Omega_c$ states within the local hidden gauge approach,''
  arXiv:1710.04231 [hep-ph], Phys. Rev. D in print.



%\bibitem{OmegacMontana}
%  G.~Monta\~na, A.~Feijoo and A.~Ramos, %\`{A}
%  %``A meson-baryon molecular interpretation for some $\Omega_c$ excited baryons,''\\
%  Eur.\ Phys.\ J.\ A {\bf 54}, no. 4, 64 (2018).
%
%\bibitem{OmegacPavao} J.~Nieves, R.~Pavao and L.~Tolos,
%  %``$\Omega _c$ excited states within a $\mathrm{SU(6)}_{\mathrm{lsf}}\times $  HQSS model,''
%  Eur.\ Phys.\ J.\ C {\bf 78}, no. 2, 114 (2018).

%\bibitem{Close}
%  F.~E.~Close,
%  \textit{``An Introduction to Quarks and Partons,''}
%  Academic Press/London 1979.%, 481p.

%\bibitem{OmegacBelle}
%  J.~Yelton {\it et al.} [Belle Collaboration],
%  %``Observation of Excited $\Omega_c$ Charmed Baryons in $e^+e^-$ Collisions,''
%  Phys.\ Rev.\ D {\bf 97}, no. 5, 051102 (2018).

%[3] M.~Karliner and J.~L.~Rosner,
%  %``Very narrow excited $\Omega_c$ baryons,'',
%  Phys.\ Rev.\ D {\bf 95}, no. 11, 114012 (2017).\\
%\vspace{3pt}
%
%[4]  H.~C.~Kim, M.~V.~Polyakov and M.~Prasza{\l}owicz,
%  %``Possibility of the existence of charmed exotica,''
%  Phys.\ Rev.\ D {\bf 96}, no. 1, 014009 (2017).\\%;
%  %Addendum: [Phys.\ Rev.\ D {\bf 96}, no. 3, 039902 (2017)].\\

\bibitem{OmegabLHCb}
R.~Aaij {\it et al.} [LHCb Collaboration],
%  %``Measurement of the mass and lifetime of the $\Omega_b^-$ baryon,''
  Phys.\ Rev.\ D {\bf 93}, 092007 (2016).

\bibitem{Omegab-Omegac}
  V.~R.~Debastiani, J.~M.~Dias, W.~H.~Liang and E.~Oset,
  %``$\boldsymbol{\Omega_b^- \to (\Xi_c^+ \, K^-) \, \pi^-}$ and the $\boldsymbol{\Omega_c}$ states,''
  arXiv:1803.03268 [hep-ph].







\end{thebibliography}

\end{document}